\documentclass[prd,twocolumn,preprintnumbers,amsmath,amssymb,nofootinbib,superscriptaddress,showpacs,showkeys]{revtex4-1}
\usepackage{amsmath,amsfonts,amssymb,amsthm,amstext,amscd,eucal,srcltx,mathrsfs}
\usepackage{epsfig,graphicx,bm}
\usepackage{epstopdf, epsf}
\usepackage{dcolumn}
\usepackage{hyperref}
\newcommand{\be}{\begin{eqnarray}}
\newcommand{\ee}{\end{eqnarray}}
\newcommand{\bra}[1]{\mbox{$\langle\, #1 \mid$}}

\newcommand{\ket}[1]{\mbox{$\mid #1\,\rangle$}}

\newcommand{\lp}{\ell_{\rm p}}
\newcommand{\mpl}{m_{\rm p}}
\newcommand{\gn}{G_{\rm N}}

\newcommand{\Rh}{R_{\rm H}}

\begin{document}
%
\title{Matter and gravitons in the gravitational collapse}
\author{Roberto Casadio}
\email{casadio@bo.infn.it}
\affiliation{Dipartimento di Fisica e Astronomia,
Alma Mater Universit\`a di Bologna,
via~Irnerio~46, 40126~Bologna, Italy}
\affiliation{I.N.F.N., Sezione di Bologna, IS FLAG
viale~B.~Pichat~6/2, I-40127 Bologna, Italy}
\author{Andrea Giugno}
\email{A.Giugno@physik.uni-muenchen.de}
\affiliation{Arnold Sommerfeld Center, Ludwig-Maximilians-Universit\"at,
Theresienstra{\ss}e 37, 80333 M\"unchen, Germany}
\author{Andrea Giusti}
\email{andrea.gusti@bo.infn.it}
\affiliation{Dipartimento di Fisica e Astronomia,
Alma Mater Universit\`a di Bologna,
via~Irnerio~46, 40126~Bologna, Italy}
\affiliation{I.N.F.N., Sezione di Bologna, IS FLAG,
viale~B.~Pichat~6/2, I-40127 Bologna, Italy}
\begin{abstract}
We consider the effects of gravitons in the collapse of baryonic matter that forms a black hole.
We first note that the effective number of (soft off-shell) gravitons that account for the (negative)
Newtonian potential energy generated by the baryons is conserved and always in agreement
with Bekenstein's area law of black holes.
Moreover, their (positive) interaction energy reproduces the expected post-Newtonian correction
and becomes of the order of the total ADM mass of the system when the size of the collapsing
object approaches its gravitational radius.
This result supports a scenario in which the gravitational collapse of regular baryonic matter
produces a corpuscular black hole without central singularity, in which both gravitons and baryons
are marginally bound and form a Bose-Einstein condensate at the critical point.
The Hawking emission of baryons and gravitons is then described by the quantum depletion
of the condensate and we show the two energy fluxes are comparable, albeit negligibly small
on astrophysical scales.
\end{abstract}
\pacs{04.70.Dy,04.70.-s,04.60.-m}
\keywords{Gravitational collapse; black holes; no singularity}
\maketitle
%
%
%
%
%
%
%
%
One of the main issues in gravity theory is to understand the formation of black holes
from the gravitational collapse of compact objects.
It is in fact a theorem in general relativity~\cite{ellis} that, provided the collapsing (massive)
matter satisfies the weak energy condition and a trapping surface appears at some
point, all the matter will eventually shrink into a space-time singularity (of infinite density).
A simple and explicit example of this behaviour was illustrated long ago by Oppenheimer and
Snyder~\cite{OS}.
However, such a singular final state of matter is clearly incompatible with the principles of quantum
physics~\footnote{A similar conclusion follows from a quantum treatment of the gravitational radius
of a quantum matter state~\cite{BEC_BH1}.}, which immediately calls for a 
search of alternative end-points of the collapse within a fully quantum description of nature.
\par
We consider here the gravitational collapse of a spherically symmetric object assuming
the validity of the Hamiltonian constraint of general relativity, that is, of total energy conservation
in Newtonian terms~\footnote{We should however remark that in Newtonian physics the total
energy of a system can be arbitrarily shifted by a constant, whereas in general relativity this
operation is not allowed.}.
The only quantum gravity ingredient we shall employ is the description of the gravitational
field in terms of ``gravitons''.
In particular, we will include the effect of (negative energy) gravitons with a wavelength of the
order of the size of the collapsing body in the total energy balance.
These soft gravitons indeed appear in the quantum representation of the Newtonian potential
by means of a coherent state coupled to the matter source~\cite{mueck,BEC_BH1}, and one might
speculate~\cite{mueckP} that they could be associated with the recently advocated breaking of
the BMS symmetry~\cite{bms} precisely induced by the presence of localised matter.
Our main result is that these (soft off-shell) gravitons satisfy Bekenstein's area law~\cite{bekenstein}
and appear to produce the expected post-Newtonian correction~\cite{weinberg} to the total energy
of the system, which becomes a major contribution to the dynamics when the gravitational radius
is approached.
At that point, a black hole should form, mostly made of such soft gravitons (in a sense that will
be clarified later on), in qualitative agreement with the corpuscular model of Refs.~\cite{dvali}.
\par
\noindent
{\em Energy balance of self-gravitating objects.\/}
%
%
%
%
%
%
%
Let us consider a simple model for a compact stellar object made of $N_{\rm B}$ identical
components, which we will call baryons for simplicity, of rest mass $\mu$ assembled in a spherically
symmetric configuration of radius $R$.
These baryons can interact gravitationally, and we assume their number does not depend on
$R$~\cite{dvaliB}.
We also neglect any emission of radiation, for simplicity, so that the total energy is conserved
and always equals the Arnowitt-Deser-Misner (ADM) mass $M$ of the system~\cite{adm}. 
Energy conservation is granted in the Newtonian description of isolated systems, but let us also recall
the same result holds in general relativity, where it is given by the Hamiltonian constraint associated
with the freedom of time reparameterization.
In an asymptotically flat space, the Hamiltonian constraint takes the form~\cite{adm,dewitt}
\be
H
\equiv
H_{\rm B}+H_{\rm G}
=
M
\ ,
\label{Hc}
\ee
where $H_{\rm B}$ and $H_{\rm G}$ respectively denote the (super)-Hamiltonian of matter and
gravity, obtained by varying the action with respect to the lapse function, and $M$ emerges
from boundary terms.
For instance, one could consider configurations of a given constant $R=R_{\rm s}$ as
representing stable stars (for which the Hamiltonian constraint leads to the Tolman-Oppenheimer-Volkov
equation), or let $R$ reduce all the way down to form a black hole of size equal
to the Schwarzschild gravitational radius~\footnote{We shall always write
the Newton constant $\gn=\lp/\mpl$ and the Planck constant $\hbar=\lp\,\mpl$.}
\be
\Rh=2\,\lp\,\frac{M}{\mpl}
\ .
\label{Rh}
\ee
\par 
If we ideally think of preparing the system when the $N_{\rm B}$ baryons are very far apart,
the total energy is simply given by the baryonic rest mass,
\be
H
=
E_{\rm B}
\equiv
\mu\,N_{\rm B}
\simeq
M
\ .
\ee
Subsequently, as the radius $R$ shrinks, the baryon energy will be decreased by the (negative)
interaction potential energy $U_{\rm BG}$ between baryons mediated by the gravitons,
and acquire kinetic energy $K_{\rm B}$, so that
\be
\label{EB}
E_{\rm B} (R)
=
M+K_{\rm B}(R) + U_{\rm BG} (R)
+U_{\rm BB}(R) 
\ ,
\ee
where we explicitly wrote the dependence of all terms on the typical size $R$ of the system,
and $U_{\rm BB}\ge 0$ is an additional (repulsive) interaction among baryons, which provides the
pressure required for a static configuration at a given finite value of $R=R_{\rm s}$ (corresponding to which
we will have $K_{\rm B}(R_{\rm s})=0$).
In a purely classical model, this would equal the total energy of the system, the ADM mass $M$,
and one would thus find the classical equation of motion
\be
K_{\rm B}(R) + U_{\rm BG} (R)+U_{\rm BB}(R)
=
0
\ . 
\label{Eb0}
\ee
\par
But baryons are quantum, and a consistent description requires we also consider quantum
features of the gravitational interaction.
For this purpose, let us start from a simple estimate of the baryon total gravitational potential
energy for a static configuration obtained from Newtonian physics, that is
\be 
U_{\rm BG} (R)
&\simeq&
N_{\rm B} \,\mu\,\phi_{\rm N}(R)
\simeq
-N_{\rm B} \,\mu\, 
\frac{\lp\, M}{\mpl\,R}
=
-\frac{M^2\,\lp}{\mpl\,R}
\ ,
\quad
\label{UbgR}
\ee
where the classical Newtonian field $\phi_{\rm N}$ satisfies the Poisson equation 
\be
\Delta\,\phi_{\rm N}
=
\lp\,\frac{M}{\mpl}\,j(r)
\ ,
\label{KG}
\ee
with a static source profile such that $\int_0^R r^2\,d r\,j(r)= 1$.
In the quantum theory, this field can be described by means of a coherent state of
(virtual) gravitons~\cite{mueck,BEC_BH1}, like a coherent state of (virtual) photons
reproduces the Coulomb field around a static charge~\cite{mueck}.
This can be easily seen from the momentum space form of Eq.~\eqref{KG},
\be
k^2\,\phi_{\rm N}(k)
=
-\frac{M}{\mpl}\,j(k)
\ ,
\ee
where $k$ is the dimensionless wave number, and expanding the graviton field
operator in the corresponding radial modes,
$\hat \phi_k \simeq (\hat g_k+\hat g_{-k}^\dagger)/\sqrt{k}$.
A coherent state is an eigenstate of the annihilation operators,
\be
\hat g_k\ket{g}=g(k)\ket{g}
\ .
\ee
In particular, we can choose
\be
g(k)\simeq -\frac{M\,j(k)}{\mpl\,k^{3/2}}
\ ,
\ee
which precisely reproduces the classical field,
\be
\bra{g}\hat\phi_k \ket{g}
\simeq
-\frac{M\,j(k)}{\mpl\,k^2}
\simeq
\phi_{\rm N}(k)
\ .
\ee
The expectation value of the graviton number is now well-approximated by 
\be
N_{\rm G}
&=&
\int k^2\,dk\,
\bra{g} \hat g_k^\dagger\,\hat g_k\ket{g}
\simeq
\frac{M^2}{\mpl^2}
\int dk\,
\frac{j^2(k)}{k}
\nonumber
\\
&\simeq&
\frac{M^2}{\mpl^2}
\sim
\frac{\Rh^2}{\lp^2}
\ ,
\label{Ngm2}
\ee
which is essentially Bekenstein's area law~\cite{bekenstein}, but holds regardless of the actual
size $R$ of the matter source.
\par
It is worth noting that, since $M$ is constant in our approximation,
the number $N_{\rm G}$ of gravitons is also conserved, like $N_{\rm B}$.
If we further write 
\be
U_{\rm BG} (R)
\simeq
N_{\rm G}\,\epsilon_{\rm G}(R)
\ ,
\ee
we immediately conclude the typical graviton energy is in fact given by~\cite{BEC_BH1,mueck} 
\be
\epsilon_{\rm G}
\simeq
-\frac{\lp}{R}\,\mpl
\ ,
\label{mL}
\ee
which is extremely small for a macroscopic source, but increases (in modulus)
for decreasing $R$.
The above relation tells us that $\epsilon_{\rm G}$ is determined by the typical
length of the quantum coherent state according to the de~Broglie relation, but is of course
negative.
This feature is in agreement with gravity contributing to the general relativistic
Hamiltonian constraint with a sign opposite to that of matter, and the non-relativistic view
of the negative Newtonian energy.
It also signals that the gravitons of a static potential are off-shell from the quantum field
theory point of view.
\par 
Since gravitons self-interact, we add the graviton interaction energy,
\be 
U_{\rm GG} (R)
\simeq
N_{\rm G}\,\epsilon_{\rm G}(R)\,\phi_{\rm N}(R) 
\simeq
N_{\rm G}\,\frac{M\,\lp^2}{R^2}
\ ,
\label{UggR}
\ee
which we note is positive and falls off with the size $R$ of the source like $1/R^2$, two properties
that characterise a typical post-Newtonian correction to the Newtonian potential~\cite{weinberg}.
Since 
\be
\left|\frac{U_{\rm GG}}{U_{\rm BG}}\right|
\simeq
\frac{\Rh}{R}
\ll 1
\ ,
\ee
this contribution is overall very small for a large star, but becomes crucial when $R\simeq\Rh$.
\par
Finally, the complete Hamiltonian constraint~\eqref{Hc} reads
\be
\label{EG}
M
&=&
E_{\rm B}
+
U_{\rm GG}
\nonumber
\\
&=&
M+K_{\rm B}(R)
+U_{\rm BB}(R)  
+U_{\rm BG} (R)
+U_{\rm GG}(R)
\ ,
\quad
\ee
which should hold for any physically acceptable value of the size $R$
of the system.
\par
\noindent
{\em Black hole configuration.\/}
Let us now consider the case that the system can contract all the way down to
$R\simeq \Rh$.
At that point we find the interesting relation
\be
U_{\rm GG} (\Rh)
\simeq
-U_{\rm BG} (\Rh)
\simeq
M
\ ,
\ee
which we can view precisely as the marginal bound condition for gravitons
of Refs.~\cite{dvali}.
It can also be recast in the form of the critical condition for a Bose-Einstein
condensate of gravitons~\cite{dvali},
\be
\alpha\,N_{\rm G}
\simeq
1
\ ,
\ee
where $\alpha\simeq \epsilon_{\rm G}^2/\mpl^2$ is the gravitational coupling for
graviton-graviton scattering.
Beside Eq.~\eqref{Ngm2}, which holds for any $R$ and can be written as
a scaling relation for the horizon radius,
\be
\Rh
\simeq
\sqrt{N_{\rm G}}\,\lp
\ ,
\ee
in the limit $R\simeq\Rh$ one also recovers the scaling relation for the effective graviton
mass~\cite{dvali}
\be
m
=
-\epsilon_{\rm G}
\simeq
\frac{\mpl}
{\sqrt{N_{\rm G}}}
\simeq
\frac{M}{N_{\rm G}}
\ .
\ee
\par
The Hamiltonian constraint in the black hole configuration now yields
\be
K_{\rm B}+U_{\rm BB}
\simeq
0
\ .
\ee
However, since we reasonably assumed $U_{\rm BB}\ge 0$, the only possible solution
allowing for the black hole formation is 
\be
K_{\rm B}(\Rh)
\simeq
U_{\rm BB}(\Rh)
\simeq
0
\ ,
\label{maxPackB}
\ee
which one can analogously view as the marginal bound condition for the baryons.
\par
Indeed one could have considered the Oppenheimer-Snyder model of collapsing dust~\cite{OS}
from the beginning, so that $U_{\rm BB}=0$ for all values of $R$, with a constant number
$N_{\rm G}$ of soft gravitons.
In this case $K_{\rm B}(\Rh)\simeq 0$ and the (quantum) matter stops collapsing. 
This of course represents a huge correction to the classical model in which the system
ends into the singularity at $R=0$.
If we took the above Eq.~\eqref{maxPackB} at face value, we could actually say that,
since any kind of matter has $U_{\rm BB}>0$ and $K_{\rm B}\ge 0$, the configuration
$R\simeq \Rh$ should not even be reached.
However, such a strong conclusion definitely requires a better analysis of all the terms in the
Hamiltonian constraint~\eqref{Hc} for $R\simeq \Rh$.
In fact, we never explicitly considered the spatial distribution of the baryons,
consequently the variable $R$ could merely represents the typical (quantum) size of the
collapsing object, rather then a sharp (classical) radius.
The simple estimates~\eqref{UbgR} and~\eqref{UggR} could then be improved by
employing a better approximation for the potential of a self-gravitating system
(for example, the harmonic~\cite{qhbh} or P\"oschl-Teller potential~\cite{mueckPT})
and one could also include effective quantum field theory corrections~\cite{donoghue}.
We leave such improvements for future investigations, and just remark that
the effective number of soft gravitons in the Newtonian potential and in the black hole
is much larger than the number of baryons.
For example, let us consider a solar mass black hole ($M = M_{\odot} \sim 10^{38} \, \mpl$)
made of neutrons ($\mu \sim 10^{-19} \, \mpl$).
The number of neutrons in the system is $N_{\rm B} = M / \mu \sim10^{57}$,
whereas the number of gravitons
$N_{\rm G}\simeq \mu\, N_{\rm B}^{2} / \mpl \sim 10^{95} \gg N_{\rm B}$,
which is again consistent with the underling hypothesis of the corpuscular model for
black holes. 
\par
\noindent
{\em Quantum depletion of gravitons and baryons.\/}
%
%
%
In the above description, we neglected any possible emission of gravitons
or baryons, but the black hole just formed should radiate according to
Hawking's law~\cite{hawking}.
In the corpuscular model, this effect is reproduced by the quantum depletion
occurring because of the graviton-graviton scatterings~\cite{dvali}.
We shall here consider the added contribution of graviton-baryon scatterings
to the emission of gravitons, and the baryon-graviton scattering for the emission
of baryons~\cite{kuhnel}.
\par
Because of the $N_{\rm B}$ baryons, the depletion law discussed in Refs.~\cite{dvali}
will become
\be
\label{depletionG}
\dot N_{\rm G}
&\simeq&
-N_{\rm G}^2\,\frac{1}{N_{\rm G}^2}\,\frac{1}{\lp\,\sqrt{N_{G}}}
-N_{\rm G}\, N_{\rm B}\,\frac{1}{N_{\rm G} ^{2}} \, \frac{1}{\lp\, \sqrt{N_{\rm G}}}
\nonumber
\\
&\simeq&
-\frac{1}{\lp \sqrt{N_{\rm G}}}
\left(1+ \frac{N_{\rm B}}{N_{\rm G}}\right)
\ ,
\ee
where in each of the two terms in the r.h.s.~of the first line, the first factor accounts
for the graviton and baryon multiplicity, the second factor is the gravitational
coupling $\alpha^2$ and the third factor comes from the typical energy $m$
of the process.
Also baryons will be scattered out of the collapsed object~\cite{kuhnel}, 
according to
\be
\label{depletionB}
\dot N_{\rm B}
&\simeq&
-N_{\rm G}\, N_{\rm B}\,\frac{1}{N_{\rm G} ^{2}} \, \frac{1}{\lp\, \sqrt{N_{\rm G}}}
\nonumber
\\
&\simeq&
-\frac{N_{\rm B}}{N_{\rm G}} \, \frac{1}{\lp\, \sqrt{N_{\rm G}}}
\ ,
\ee
where we assumed the typical baryon energy will again be given by the typical
graviton's energy, as predicted by Hawking.
It is then important to notice that baryons with such a small energy can be emitted
because Eq.~\eqref{maxPackB} holds after the black hole is formed.
Looking at the above emission rates, although $\dot N_{\rm G}\gg \dot N_{\rm B}$,
it is clear that the corresponding energy fluxes are of the same magnitude.
\be
m\,\dot N_{\rm G}
\simeq
\frac{M}{\lp\,N_{\rm G}^{3/2}}
\simeq
\mu\,\dot N_{\rm B}
\ .
\ee
This implies that both are indeed practically irrelevant for an astrophysical object with
$M\gg \mpl$ (and thus $N_{\rm G}\gg N_{\rm B}\gg 1$),
like one already expected from the standard expression of the Hawking effect.
%
%
%
%
%
%
\par
\noindent
{\em Concluding remarks.\/}
%
%
%
%
We have shown, in rather general terms, that including the effect of soft
gravitons in the description of the gravitational collapse of a compact object
naturally leads to the expected post-Newtonian correction to the energy
of the system and to the possible formation of a corpuscular black hole
mostly made of gravitons.
By this statement we precisely mean that the number of gravitons
$N_{\rm G}\gg N_{\rm B}$ and their typical effective mass $m=-\epsilon_{\rm G}$
is such that $M\simeq N_{\rm G}\,m$~\cite{dvali}.
In astrophysical situations, where the Hawking radiation, here described
as a depletion effect, is negligible, this state should represent the end-point
of the collapse, with no central singularity at all.
\par
Of course, our conclusion could be further refined by considering 
improved approximations for the various energy terms that appear in
the Hamiltonian constraint, like we commented previously.
Moreover, we did not solve for any specific dynamics, and it is thus
possible that different energy terms will appear in a fully time-dependent
analysis.
Even if all the terms remained of the same functional form, numerical coefficients
might very likely differ from those we employed, and this is the main reason
we did not show many exact equalities in the present analysis.
Of course, it would be extremely interesting to derive the macroscopic
dynamics from the microscopic (quantum field theory) description of
(corpuscular) black hole formation from graviton-graviton (and graviton-baryon)
scatterings~\cite{dlust,veneziano,cx}.
All that said, our findings still suggest to consider the possibility that the
end-point of the gravitational collapse, or physical black holes, are quite different
objects from those described by classical general relativity, and their quantum
properties might therefore differ from the usual ones obtained from quantum
field theory on classical black hole backgrounds.
\begin{acknowledgments}
We would like to thank R.~da~Rocha, G.~Dvali, A.~Y.~Kamenshchik, O.~Micu,
W.~M\"uck and G.~Venturi for valuable comments and discussions.  
\end{acknowledgments}
\end{document}